\documentclass{article}
\usepackage[hidelinks]{hyperref}
\usepackage{amsthm,amsmath,amssymb}
\usepackage{rotating}
\usepackage{titlesec}
\usepackage{mathrsfs}
\usepackage{multirow}
\RequirePackage{enumitem}
\RequirePackage{mathtools}
\usepackage[doublespacing]{setspace}
\usepackage{geometry}
\usepackage{natbib}
\usepackage{color, colortbl}
\definecolor{Gray}{gray}{0.9}

\numberwithin{equation}{section}
\theoremstyle{plain}

\titlelabel{\thetitle.\quad}

\begin{document}

\title{\textbf{Optimal Bayesian stepped-wedge cluster randomised trial designs for binary outcome data}}
\author{\textbf{Laura Etfer\textsuperscript{1}*, James M.S. Wason\textsuperscript{1}, Michael J. Grayling\textsuperscript{2}}\\
\small 1. Biostatistics Research Group, Newcastle University,\\
\small 2. Johnson \& Johnson Innovative Medicine\\
\small Corresponding author: Laura Etfer; Address: Ridley Building 1, Lovers Lane,\\
\small Newcastle upon Tyne NE1 7RU, UK; E-mail address: l.k.etfer@newcastle.ac.uk}
\date{}
\maketitle

\section*{Abstract}
\textbf{Introduction:} A number of recent articles have investigated the optimal design of stepped-wedge cluster randomised trials. Each of these has focused on the case of normally distributed outcomes, analysed using a linear mixed model. This is unfortunate as many stepped-wedge trials have a binary primary outcome. We therefore demonstrate how Bayesian methods can be used to find optimised designs for binary outcomes. \\
\textbf{Methods:} Under a generalised estimating equation analysis approach, approximate design theory is used to determine Bayesian D-optimal designs. For two examples, considering simple exchangeable and exponential decay correlation structures, we compare the efficiency of identified optimal designs to balanced stepped-wedge designs and corresponding stepped-wedge designs determined by optimising using a normal approximation approach.\\
\textbf{Results:} The dependence of the Bayesian D-optimal designs on the assumed correlation structure is explored; for the considered settings, smaller decay in the correlation between outcomes across time periods, along with larger values of the intra-cluster correlation, leads to designs closer to a balanced design being optimal. Unlike for normal data, it is shown that the optimal design need not be centro-symmetric in the binary outcome case. The efficiency of the Bayesian D-optimal design relative to a balance design can be large, but situations are demonstrated in which the advantages are small. Similarly, the optimal design from a normal approximation approach is often not much less efficient than the Bayesian D-optimal design.\\
\textbf{Conclusions:} Bayesian D-optimal designs can be readily identified for stepped-wedge cluster randomised trials with binary outcome data. In certain circumstances, principally ones with strong time period effects, they will indicate that a design unlikely to have been identified by previous methods may be substantially more efficient. However, they require a larger number of assumptions than existing optimal designs, and in many situations existing theory under a normal approximation will provide an easier means of identifying an efficient design for binary outcome data.\\
\textbf{Keywords:} Approximate design theory; Bayesian design; Cluster crossover; Efficient design; Generalised estimating equation; Optimal design; Robust design.

\section{Introduction} \label{sec:intro}

A stepped-wedge (SW) cluster randomised trial (CRT) is a form of crossover trial in which clusters (i.e., groups of participants) are randomised to sequences that transition from a control to an intervention condition in a unidirectional manner.
SW-CRTs are an increasingly popular type of study design, owing to several considerations, with a desire to provide the intervention to all clusters in the trial being cited as the most common reason, followed by logistical and methodological reasons \citep{grayling2017}.
Indeed, there are now a large number of practical applications of SW-CRT designs within the literature, alongside an extensive methodological research base.
Within the available methodological literature, there are in particular a number of papers concerning efficient and optimal design. 

An optimal design is, in general, one which allows us to produce a higher quality of results according to some measure, e.g., by minimising the variance of the estimated treatment effect or by maximising the power of the study.
Literature concerning optimal SW-CRT design started with work by \citet{lawrie2015}, who derived the optimal design for SW-CRTs under analysis by the \citet{hussey2007} model.
\citet{li2018b} later extended this result to cohort SW-CRT designs.
\citet{thompson2017} investigated the optimal form of SW-CRT designs given an equal allocation of clusters to sequences, assuming normally distributed data.
\citet{zhan2018}, by contrast, looked at optimal design when clusters are allowed to not be sampled from at certain stages of the trial. 
More recently, \citet{singh2024} used Bayesian methods to find optimised SW-CRT designs that address the dependency of SW-CRTs on the intra-cluster correlation (ICC), by placing priors on the ICC value.

A common assumption in the literature on optimal SW-CRT design is that the data to be accrued is normally distributed.
In practice, though, it is common for SW-CRTs to accrue data with other distributional forms as their primary endpoint.
This is reflected in the wider literature on SW-CRT design through a number of papers.
For example, looking at the design and analysis of SW-CRTs with binary outcomes, we have such works as \citet{barker2017}, who suggested a generalised linear mixed model (GLMM) analysis is best for binary outcome SW-CRTs with a small number of clusters.
Furthermore, the work of \citet{harrison2021} proposed power calculation methods for binary outcome SW-CRTs assuming a generalised estimating equation (GEE) analysis, while \citet{zhou2018} proposed a method for power calculation with binary data via maximum likelihood estimation.
Sample size calculation for SW-CRTs with binary outcomes has also been investigated by \citet{wang2021} and \citet{li2018a}, who both proposed a method based on GEE analysis.
Finally, \citet{ford2020} investigated how using a GEE analysis for small sample SW-CRTs with binary outcomes can maintain the validity of inference compared to a GLMM.

While these works provide valuable results for the design and analysis of SW-CRTs, they leave open the question of how to optimally choose sequences in SW-CRTs with binary outcomes.
It is this problem we focus on in this paper.
For this task, we leverage techniques previously employed by, e.g., \citet{singh2016}, who describe a Bayesian approach to determining optimal treatment sequences in an individually randomised cross-over trial for three scenarios involving non-normal data.
Specifically, an important consideration for analysis of binary data in SW-CRTs is that the variance of the treatment effect estimator is dependent on several unknown model parameters.
Whilst one could attempt to surmount this problem by assuming values for these parameters, and then using techniques similar to those for the optimal design of SW-CRTs with normal data, a principal concern would then be that the resulting optimal design is only locally optimal to the assumed model parameter values.
Therefore, to overcome this problem in the employed Bayesian approach, prior distributions for the required model parameters are introduced; the treatment effect variance for different values of the model parameters is then computed and the priors are incorporated into an objective function as weighting factors.
We assume a GEE based-approach to modelling the marginal cluster-period means, with the correlation between measurements across time periods within a cluster modelled through `simple exchangeable' or `exponential decay' working correlation structures.

Ultimately, the key questions we seek to answer are what the form of the optimal design for binary data may look like in a SW-CRT, how this optimal design varies depending on the underlying parameters of the assumed model, and whether the optimal design computed under a normal approximation approach is similar to that identified using our proposed, more complex, optimal design procedure.
Before we address these questions, via several examples, we first describe the proposed framework for optimal design.
We conclude with discussion of the limitations of our work and some suggestions for future research.

\section{Methods} \label{sec:methods}

\subsection{Model} \label{subsec:model}

We consider a cross-sectional SW-CRT such that there are $I$ clusters, $J$ time periods, and $K$ participants per cluster per time period from which outcomes are accrued.
Thus, outcomes $Y_{ijk} \in\{0, 1\}$, $i \in \{1, \dots, I\}$, $j \in \{1, \dots, J\}$, $k \in \{1, \dots, K\}$ are collected.
As in, e.g., \citet{li2022}, the marginal mean model is assumed to be $g(\mu_{ijk}) = \beta_j + X_{ij}\delta$, where $\mu_{ijk}$ is the mean of $Y_{ijk}$, $g(\cdot)$ is the logit link function, $\beta_j$ is the $j$th period effect, $X_{ij} \in \{0, 1\}$ is the intervention indicator, and $\delta$ is the time-adjusted average intervention effect on the link function scale.
Next, let the sum of the cluster period outcomes be $Y_{ij} = \sum_{k = 1}^K Y_{ijk}$, the cluster-period means be $\bar{Y}_{ij} = Y_{ij}/K$, and set $\bar{\boldsymbol{Y}}_i = (\bar{Y}_{i1}, \dots, \bar{Y}_{iJ})^\top$.
We let the mean of $\bar{\boldsymbol{Y}}_i$ be $\boldsymbol{\mu}_i = (\mu_{i1}, \dots, \mu_{iJ})^\top$, where as we assume a binary outcome $\mu_{ij} = \mathbb{E}(Y_{ij})/K$ is the prevalence in the $(i, j)$th cluster-period.
Using the marginal model above, we have $g(\mu_{ij}) = \beta_j + X_{ij}\delta$.
Setting $\boldsymbol{\theta} = (\beta_1, \dots, \beta_J, \delta)^\top$, the GEEs for $\boldsymbol{\theta}$ are \citep{liang1986}
$$ \sum_{i = 1}^{I} \boldsymbol{D}_i^\top \boldsymbol{V}_i^{-1}(\bar{\boldsymbol{Y}}_i - \boldsymbol{\mu}_i) = \boldsymbol{0}, $$
where $\boldsymbol{D}_i = \partial\boldsymbol{\mu}_i/\partial\boldsymbol{\theta}^\top$ and $\boldsymbol{V}_i = \text{Cov}(\bar{\boldsymbol{Y}}_i)$ is a working covariance structure for the outcome vector $\bar{\boldsymbol{Y}}_i$.

In computing optimal designs, we assume $\boldsymbol{V}_i$ is either simple exchangeable or exponential decay in form.
Using \citet{li2022}, in the simple exchangeable case, the diagonal elements of $\boldsymbol{V}_i$ are
$$ \text{Var}(\bar{Y}_{ij}) = \frac{\nu_{ij}}{n} \{1 + (n - 1)\alpha_0\}, $$
where $\nu_{ij} = \mu_{ij}(1 - \mu_{ij})$ is the binomial variance.
Off-diagonal elements are
$$ \text{Cov}(\bar{Y}_{ij}, \bar{Y}_{il}) = \sqrt{\nu_{ij}\nu_{il}}\alpha_0. $$
I.e., a constant correlation $\alpha_0$ is assumed between any two individual outcomes from the same cluster, regardless of the time periods they are from.

In the exponential decay case, the diagonal elements are as above, whilst off-diagonal elements are instead given by
$$ \text{Cov}(\bar{Y}_{ij}, \bar{Y}_{il}) = \sqrt{\nu_{ij}\nu_{il}}\alpha_0\rho^{|j - l|}. $$
That is, the correlation between two outcomes measured in the $j$th and $l$th periods is $\alpha_0\rho^{|j - l|}$.
Note that this reduces to the simple exchangeable structure in the case $\rho = 1$.

\subsection{Approximate design} \label{subsec:approx}

We utilise approximate design theory \citep{laska1983,kushner1997,kushner1998} to find optimal SW-CRT designs.
To achieve this, first note that asymptotically the GEE estimator $\boldsymbol{\hat{\theta}}$ is approximately multivariate normal with mean $\boldsymbol{\theta}$ and covariance \citep{zeger1988}
$$ \text{Var}(\boldsymbol{\hat{\theta}}) = \left( \sum_{i = 1}^{I} \boldsymbol{D}_i^\top \boldsymbol{V}_i^{-1} \boldsymbol{D}_i \right)^{-1}. $$
Next, observe that the values of $\boldsymbol{D}_i$ and $\boldsymbol{V}_i$, by the assumed model, depend only on the treatment sequence to which the cluster was allocated.
Thus, we can re-write $\text{Var}(\boldsymbol{\hat{\theta}})$ as follows.
Let $\boldsymbol{D}_s$ and $\boldsymbol{V}_s$ be the matrices associated with the sequence $s = 2, \dots, J$ that switches to the intervention condition in period $s$.
Then let $p_s$ be the proportion of clusters allocated to sequence $s$, with $\sum_{s = 2}^J p_s = 1$.
We then have
$$ \text{Var}(\boldsymbol{\hat{\theta}}) = \frac{1}{I}\left( \sum_{s = 2}^J p_s \boldsymbol{D}_s^\top \boldsymbol{V}_s^{-1} \boldsymbol{D}_s \right)^{-1}. $$
In approximate design theory, a design is then specified by $\boldsymbol{p} = (p_2, \dots, p_J)$.
Observe also that $\text{Var}(\boldsymbol{\hat{\theta}})$ is inversely proportional to $I$, and thus the optimal sequences will not be dependent on $I$.

\subsection{Optimal design} \label{subsec:optimal}

As our principal interest is in estimating the treatment effect, we focus on $\text{Var}(\hat{\delta}) = (0, \dots, 0, 1)\text{Var}(\boldsymbol{\hat{\theta}})(0, \dots, 0, 1)^\top$.
Then, the design minimising the criterion
$$ \Lambda(\boldsymbol{p}, \boldsymbol{\theta}, \alpha_0, \rho) = \log \{\text{Var}(\hat{\tau})\}, $$
is known as the $D_A$-optimal design \cite{atkinson2007}, where the dependence of the right-hand side on $\boldsymbol{p}$, $\boldsymbol{\theta}$, $\alpha_0$, and $\rho$ is left implicit.

Given our analysis assumptions, as was noted earlier, the obtained optimal design is only locally optimal. 
Thus, to obtain $D_A$-optimal designs robust to uncertainties in the parameters we use a Bayesian approach.
This method has been used by \citet{singh2016} for cross-over trials and is based on works by \citet{chaloner1989} and \citet{dror2006} for logistic regression and for the block designs by \citet{woods2011}. 
Specifically, the $D_A$-optimal Bayesian SW-CRT design is the design that minimises
\begin{equation}\label{eq:objfn}
\Psi(\boldsymbol{p}, \boldsymbol{\Theta}, \alpha_0, \rho) = \int_{\boldsymbol{\Theta}} \Lambda(\boldsymbol{p}, \boldsymbol{\theta}, \alpha_0, \rho) \ dF(\boldsymbol{\theta}),
\end{equation}
where $\Theta \subset \mathbb{R}^{J + 1}$ is the parameter space of parameter vector $\boldsymbol{\theta}$ and $F(\boldsymbol{\theta})$ is a proper prior distribution for $\boldsymbol{\theta}$ \citep{pettersson2005}.

Note that by the above, in our examples no prior distributions are assigned to the correlation parameters $\alpha_0$ and $\rho$; designs are obtained only for some fixed values chosen for these parameters.
Our approach could be readily extended to place priors on these parameters; as demonstrated by \cite{singh2024} this may be helpful for $\alpha_0$.
By contrast, placing anything but a vague prior on $\rho$ may be challenging in practice as there is often little known about this parameter at the design stage.

In our computations, we assume a prior for $\beta_j$ of the form $U(\beta_{lj}, \beta_{uj})$, and similarly for $\delta$ as $U(\delta_l, \delta_u)$.
We then take the multi-dimensional prior for $\boldsymbol{\theta}$ as the Cartesian product of these uniform priors.
Further information on how $\delta_l$, $\delta_u$, the $\beta_{lj}$ and $\beta_{uj}$ are specified is given below.

The minimisation of the objective function in Equation~(\ref{eq:objfn}) with respect to $\boldsymbol{p}$ requires high-dimensional integral calculation.
As in \citet{singh2016}, we leverage Latin Hypercube Sampling (LHS) and an interior-point optimization algorithm to derive the solution of the above optimisation problem. 

\subsection{Design comparison} \label{subsec:comparison}

To assess the performance of an optimal design, $\boldsymbol{p}_\text{optimal}$ say, we compare its efficiency relative to a balanced design and an optimal design computed under a simpler normal approximation approach using the results of \cite{lawrie2015}.
That is, we compare to
\begin{align*}
  \boldsymbol{p}_\text{balanced} &= (1, 1, \dots, 1)/(J - 1), \\
  \boldsymbol{p}_\text{lawrie}   &= (p_\text{outer}, p_\text{inner}, p_\text{inner}, \dots, p_\text{inner}, p_\text{outer}), \\
  p_\text{outer}                 &= \frac{1 + \alpha_0(3K - 1)}{2\{1 + \alpha_0(JK - 1)\}}, \\
  p_\text{inner}                 &= \frac{K\alpha_0}{1 + \alpha_0(JK - 1)}.
\end{align*}

To measure the efficiency advantage of the optimal design, we assess the percentage of additional clusters that would be required by a design using sequence proportions $\boldsymbol{p}_\text{balanced}$ or $\boldsymbol{p}_\text{lawrie}$, to achieve the same value of the objective function as a design using proportions $\boldsymbol{p}_\text{optimal}$.
Using the fact that $\text{Var}(\boldsymbol{\hat{\theta}})$ is inversely proportional to $I$, this can be computed for the balanced design as
$$ \frac{\exp\{\Psi(\boldsymbol{p}_\text{balanced}, \boldsymbol{\Theta}, \alpha_0, \rho)\}}{\exp\{\Psi(\boldsymbol{p}_\text{optimal}, \boldsymbol{\Theta}, \alpha_0, \rho)\}}, $$
and simply for the design from \cite{lawrie2015}.

\subsection{Examples} \label{subsec:examples}

Code to reproduce all results is available from \url{https://github.com/lauraetfer/optimal_SWCRT}.

\subsubsection{Washington State EPT trial} \label{subsubsec:wash}

The Washington State EPT trial \citep{golden2015} was a SW-CRT used to determine the effect of expedited partner therapy on chlamydia and gonorrhoea rates.
It utilised a SW-CRT design with four sequences ($J = 5$) and $I = 22$ clusters.
\citet{li2022} provide estimated values for the model parameters under the simple exchangeable and exponential decay structures; these are replicated in Table~\ref{tab:wash}.
We compute the Bayesian optimal design for these model parameters, setting the required uniform priors for the components of $\boldsymbol{\theta}$ as the 95\% Wald confidence interval using the point estimates and standard errors shown.
We assume $K = 305$, reflecting the average cluster-period size of the trial.

\begin{table}[htbp]
\centering
\caption{Parameter estimates of marginal mean and correlation parameters for the Washington State EPT trial, as reported by \cite{li2022}. Results are shown for the simple exchangeable and exponential decay correlation structures. Standard error estimates are shown in parentheses.}
\begin{tabular}{rrr}
\hline
 Parameter & Simple exchangeable & Exponential decay \\
\hline
 $\beta_1$ &      -2.444 (0.091) &    -2.437 (0.095) \\
 $\beta_2$ &      -2.454 (0.091) &    -2.444 (0.089) \\
 $\beta_3$ &      -2.535 (0.094) &    -2.508 (0.100) \\
 $\beta_4$ &      -2.609 (0.106) &    -2.613 (0.115) \\
 $\beta_5$ &      -2.537 (0.145) &    -2.552 (0.131) \\
  $\delta$ &      -0.141 (0.092) &    -0.124 (0.087) \\
$\alpha_0$ &     0.0051 (0.0016) &   0.0070 (0.0039) \\
    $\rho$ &                  -- &   0.7157 (0.2962) \\
\hline
\end{tabular}
\label{tab:wash}
\end{table}

\subsubsection{Hypothetical design scenario} \label{subsubsec:hyp}

To explore how the design assumptions influence the resulting optimal design in further detail, we consider a wide-range of parameter combinations, motivating our assumptions by approaches previously taken in \cite{barker2017} and \cite{li2022}.
Specifically, we assume $J = 9$, reflecting a common number of time periods in practice, and assume $\delta = log(2.25)$ as in \cite{barker2017}.
For the time trends, we take a generalised version of the assumptions in \cite{li2022}, assuming $\beta_1 = 0.1$, with $\beta_j = \beta_{j - 1} + ab^{j - 1}$ for $j = 2, \dots, J$.
We then consider $a \in \{-0.5, -0.4, \dots, 0.5\}$ and $b \in \{0, 0.1, \dots, 1\}$.
Additionally, we examine settings in which $K \in \{10, 50, 100\}$, $\alpha_0 \in \{0.01, 0.02, \dots, 0.1\}$, and $\rho \in \{0.5, 0.55, \dots, 1\}$.
To set the multi-dimensional prior assumption for particular $\boldsymbol{\theta}$ (implied by specific values of $a$ and $b$), we compute $\text{Var}(\boldsymbol{\theta})$ and then set the lower and upper bounds of the uniform priors for the components of $\boldsymbol{\theta}$ using the 80\% Wald confidence intervals for this computed covariance matrix.
This approach is designed to reflect having constructed the multi-dimensional prior from a completed pilot study that leveraged an equivalent SW-CRT design, though we acknowledge the choice of 80\% confidence interval is somewhat arbitrary.

\section{Results} \label{sec:results}

\subsection{Washington State EPT trial} \label{subsec:wash}

For the assumptions relating to the simple exchangeable model, the optimal allocation of clusters to sequences for the Washington State EPT trial was found to be $\boldsymbol{p}_\text{se} = (0.3493, 0.1761, 0.1780, 0.2966)$.
For the exponential decay model, this was similar, instead being $\boldsymbol{p}_\text{ep} = (0.3401, 0.1882, 0.1817, 0.2900)$.
These designs are shown visually in the Supplementary Materials in Figure~S1.
Thus, in both instances the largest proportion of clusters should be assigned to the sequence that first switches to the intervention condition in time period 2.
Note also that in both cases the optimal design is not centro-symmetric; this contrasts the balanced design, $\boldsymbol{p}_\text{balanced} = (0.25, 0.25, 0.25, 0.25)$, and optimal allocations from a normal approximation under the simple exchangeable model provided by \cite{lawrie2015}, $\boldsymbol{p}_\text{lawrie} = (0.3227, 0.1773, 0.1773, 0.3227)$.
With $\boldsymbol{p}_\text{lawrie}$ similar to both $\boldsymbol{p}_\text{se}$ and $\boldsymbol{p}_\text{ep}$, this design is only minorly less efficient ($< 1\%$) than the optimal designs for either model.

\subsection{Influence of time trend on optimal design} \label{subsec:time}

\begin{figure}[htbp]
    \centering
    \includegraphics[width=\textwidth]{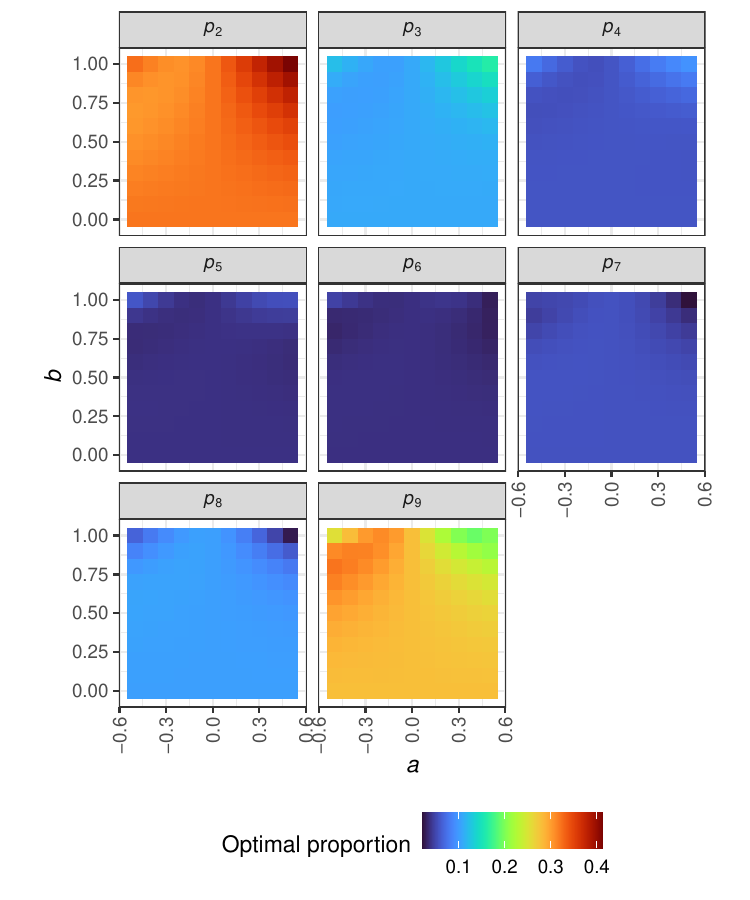}
    \caption{Optimal values of $p_2, \dots, p_9$ in the case where $K = 50$, $\alpha_0 = 0.05$, and $\rho = 0.5$ (exponential decay), as a function of time trend parameters $a$ and $b$.}
    \label{fig:time_exp}
\end{figure}

\begin{figure}[htbp]
    \centering
    \includegraphics[width=\textwidth]{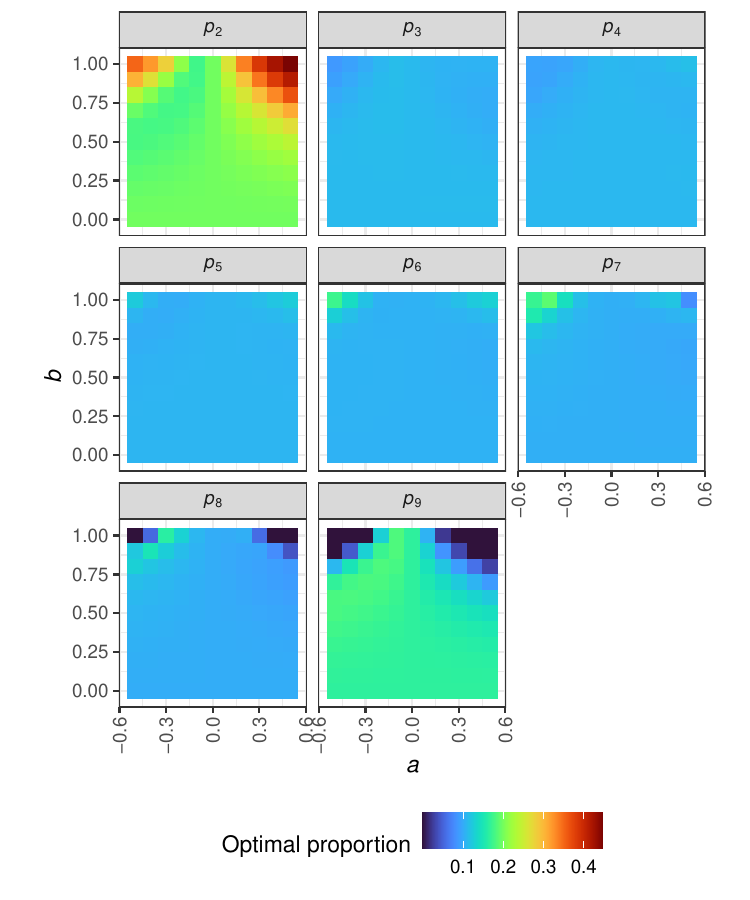}
    \caption{Optimal values of $p_2, \dots, p_9$ in the case where $K = 50$, $\alpha_0 = 0.05$, and $\rho = 1$ (simple exchangeable), as a function of time trend parameters $a$ and $b$.}
    \label{fig:time_sim}
\end{figure}

Figures~\ref{fig:time_exp}-\ref{fig:time_sim} examine how the assumptions regarding the time trend (i.e., $\beta_1, \dots, \beta_J$) influence the optimal sequences.
Specifically, Figure~\ref{fig:time_exp} assumes $K = 50$, $\alpha = 0.05$, and $\rho = 0.5$ (i.e., an exponential decay model), and then presents the optimal sequences as a function of $(a, b) \in \{-0.5, -0.4, \dots, 0.5\} \times \{0, 0.1, \dots, 1\}$.
Figure~\ref{fig:time_sim} gives the equivalent findings, but assuming $\rho = 1$ (i.e., the simple exchangeable model).
In the Supplementary Materials, additional (similar) findings are given for the case $\alpha_0 = 0.01$.

We observe that across both figures, $p_2$ and $p_9$ generally have the largest allocations.
This is particularly true when $b$ is not large.
However, as $b$ approaches 1 the value of $p_9$ is observed to decrease rapidly, particularly when $a$ is large.
That is, for strong time trend effects, the sequence that first allocates the intervention in time period 9 provides lower utility, with this especially true when the assumed time trend is positive.
Similar, but less extreme observations are observed for the influence of $a$ and $b$ on the values of $p_3$ and $p_8$.
Importantly, we observe again that the optimal design need not be centro-symmetric, with $p_2$ larger than $p_9$, and $p_3$ larger than $p_8$.

Observe also that in Figure~\ref{fig:time_exp}, $p_3$ and $p_8$ are generally larger than $p_4$-$p_7$.
However, in Figure~\ref{fig:time_sim}, these is more uniformity in the values of $p_3$-$p_8$.
Thus, for the simple exchangeable model, when the assumed time trend is weak the optimal design is similar to that provided by \cite{lawrie2015}, as may be expected.

\subsection{Influence of covariance structure on optimal design} \label{subsec:cov}

\begin{figure}[htbp]
    \centering
    \includegraphics[width=\textwidth]{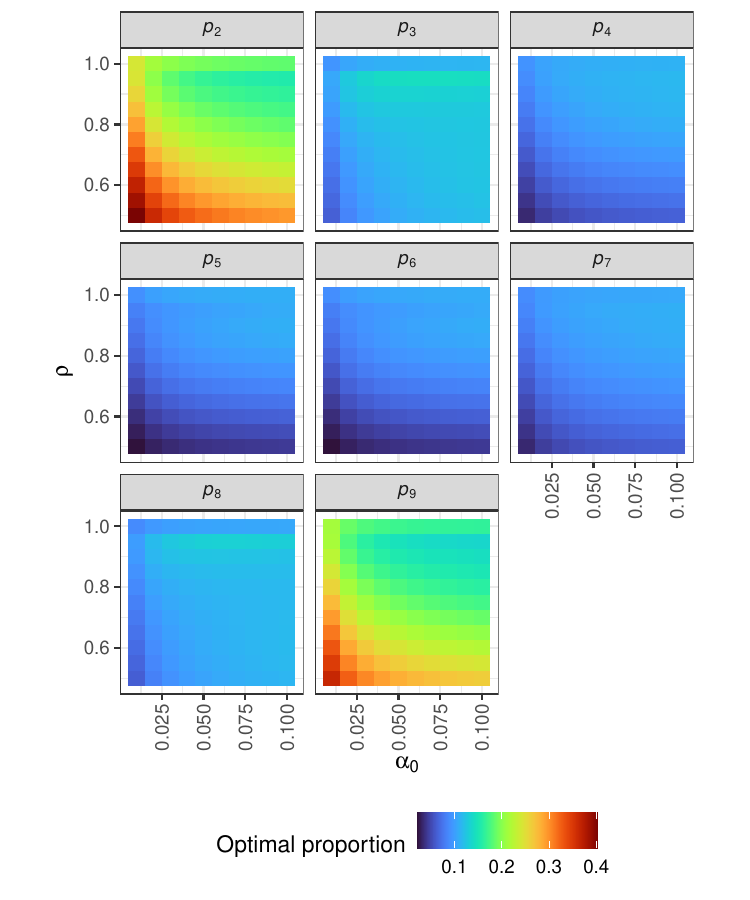}
    \caption{Optimal values of $p_2, \dots, p_9$ in the case where $K = 50$, $a = -0.1$, and $b = 0.5$, as a function of covariance structure parameters $\alpha_0$ and $\rho$.}
    \label{fig:cov_weak}
\end{figure}

\begin{figure}[htbp]
    \centering
    \includegraphics[width=\textwidth]{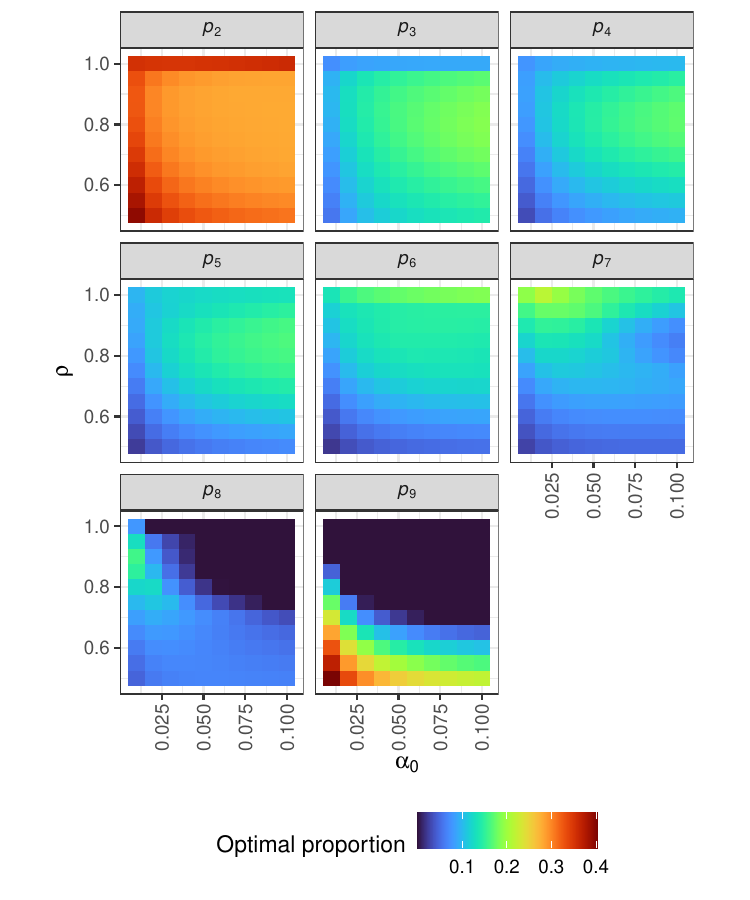}
    \caption{Optimal values of $p_2, \dots, p_9$ in the case where $K = 50$, $a = -0.5$, and $b = 1$, as a function of covariance structure parameters $\alpha_0$ and $\rho$.}
    \label{fig:cov_strong}
\end{figure}

Figures~\ref{fig:cov_weak}-\ref{fig:cov_strong} examine how the assumptions regarding the covariance structure (i.e., $\alpha_0$ and $\rho$) influence the optimal sequences.
Specifically, Figure~\ref{fig:cov_weak} assumes $K = 50$, $a = -0.1$, and $b = 0.5$ (i.e., a weak negative time trend), and then presents the optimal sequences as a function of $(\alpha_0, \rho) \in \{0.01, 0.02, \dots, 0.1\} \times \{0.5, 0.55, \dots, 1\}$.
Figure~\ref{fig:cov_strong} gives the equivalent findings, but assuming $a = -0.5$, and $b = 1$ (i.e., a strong negative time trend).

In Figure~\ref{fig:cov_weak}, we observe a shift towards a more balanced design being optimal when $\alpha_0$ and $\rho$ are larger.
By contrast, small $\alpha_0$ and/or $\rho$ results in far greater allocation to sequences 2 and 9 being optimal.
Interestingly, the values of $p_4$-$p_7$ are less impacted by the assumptions regarding the correlation structure.

In Figure~\ref{fig:cov_strong} a more extreme pattern is observed.
Whilst sequence 2 is always given a large allocation, sequences 8 and 9 are effectively not utilised when $\alpha_0$ and $\rho$ are large.
There is also in this case greater variation in the values of $p_4$-$p_7$.
This tells us that in the case where a strong time trend is assumed, the optimal design is more likely to not resemble previously identified optimal designs.

\subsection{Efficiency advantage of optimal design} \label{subsec:eff}

\begin{figure}[htbp]
    \centering
    \includegraphics[width=\textwidth]{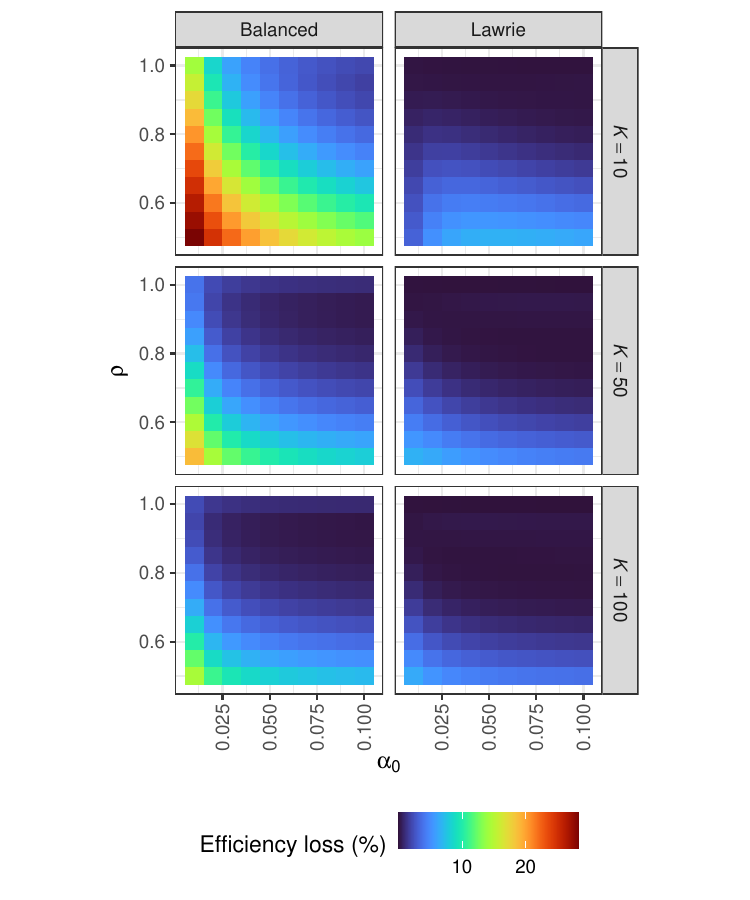}
    \caption{Efficiency loss of balanced and Lawrie designs, relative to the optimal design, in the case where $K = 50$, $a = -0.1$, and $b = 0.5$, as a function of covariance structure parameters $\alpha_0$ and $\rho$.}
    \label{fig:eff_weak}
\end{figure}

\begin{figure}[htbp]
    \centering
    \includegraphics[width=\textwidth]{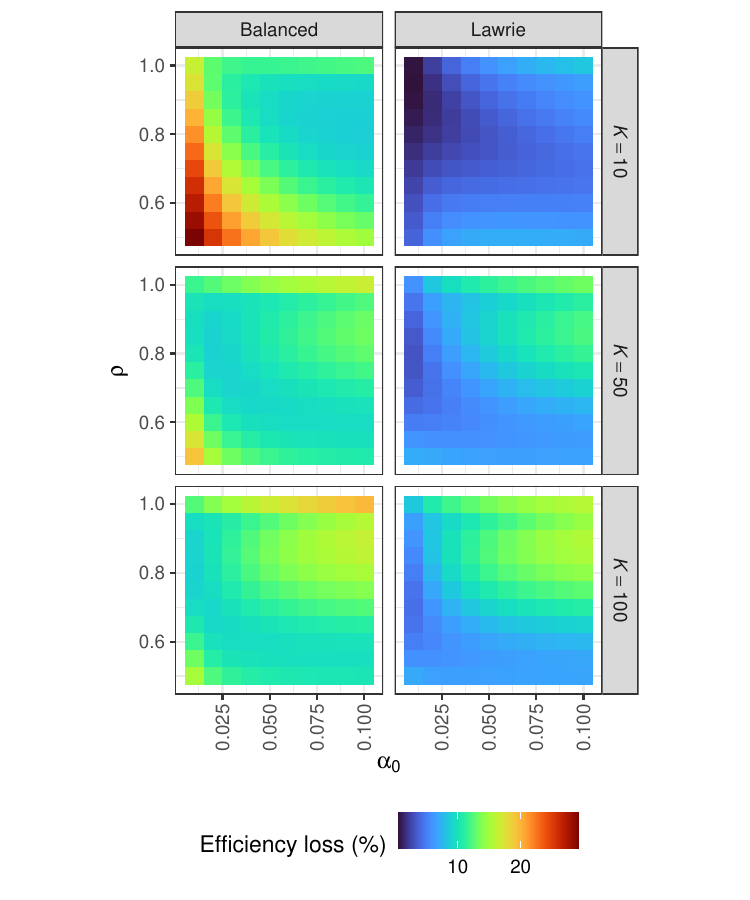}
    \caption{Efficiency loss of balanced and Lawrie designs, relative to the optimal design, in the case where $K = 50$, $a = -0.5$, and $b = 1$, as a function of covariance structure parameters $\alpha_0$ and $\rho$.}
    \label{fig:eff_strong}
\end{figure}

The previous subsections outline when the optimal design may differ greatly from optimal designs identified under methods developed assuming normally distributed data presented previously.
A key question though is whether when there are such differences, does the resulting optimal design provide a marked improvement in efficiency over the simpler historical designs.
Figures~\ref{fig:eff_weak} and~\ref{fig:eff_strong} examine this by comparing the performance of the optimal allocation of clusters to a balanced design and to the allocation given by \citet{lawrie2015}. 
Specifically, they show the percentage increase in clusters required by balanced or `Lawrie' designs to achieve the same objective function value as the Bayesian optimal design.
Both figures relate to the case where $K \in \{10, 50, 100\}$, and both show results for $(\alpha_0, \rho) \in \{0.01, 0.02, \dots, 0.1\} \times \{0.5, 0.55, \dots, 1\}$.
They differ in that Figures~\ref{fig:eff_weak} relates to a weak decreasing time trend ($a = -0.1$, $b = 0.5$) and Figures~\ref{fig:eff_strong} to a strong decreasing time trend ($a = -0.5$, $b = 1$).

Importantly in Figures~\ref{fig:eff_weak} we observe, as may be expected, that there is generally only a small efficiency gain from the optimal design compared to the design provided by \citet{lawrie2015}.
The efficiency gain relative to a balanced design can be large, particularly when $K = 10$ and $\alpha_0$ and $\rho$ are small.
by contrast, Figure~\ref{fig:eff_strong} points to a greater potential for the optimal design to provide notable efficiency advantages compared to the \cite{lawrie2015} design.
Specifically, an efficiency loss of approximately 15\% is observed for a range of $(\alpha_0, \rho)$ assumptions when $K \in \{50, 100\}$.

\section{Discussion}

This article expanded on the design literature for SW-CRTs that use binary outcome variables.
Optimal allocation of clusters to sequences was investigated, with a Bayesian approach being utilised due to the presence of unknown model parameters in the correlation structure.
Allocations were obtained by minimising an objective function interpretable as a weighted treatment effect variance.
The performance of the optimal design was then compared to a traditional balanced design, as well as an optimal design assuming outcomes were normally distributed under a simple exchangeable correlation structure. 

We found that the optimal allocations can be sensitive to the assumption regarding the correlation parameters, with this more true when strong time trend effects are assumed.
Furthermore, it was observed that in this setting the optimal design need not be centro-symmetric.
Critically, though, in many instances the sequence allocations provided by \cite{lawrie2015} result in a design almost as efficient as the more complex Bayesian optimal designs.
This means in practice one may achieve the majority of possible efficiency gains using the simple closed form optimal design under a normal approximation.
Only when very strong time trends are anticipated would this likely not be the case.
In this instance, up to $\sim 15\%$ more clusters may be required by the design from \cite{lawrie2015} to achieve the same performance as the considered Bayesian optimal designs.

We conclude by acknowledging some limitations to our work and avenues for future research.
First, the effect of misspecification when calculating the optimal design was not investigated.
In particular, we did not examine sensitivity to misspecification of $\alpha_0$ or $\rho$ on the performance of the optimal design (i.e., when and whether an optimal design developed assuming particular values $(\alpha_0, \rho)$ would perform poorly if the true model parameters are instead $(\alpha_0', \rho')$).
In practice, investigating this may be an important step to ascertaining whether to leverage a given optimal design.
Similarly, while we expect there to be a degree of robustness to the design due to the Bayesian averaging over the space where we believe the true values of parameters to lie, the effect of the choice of priors remains an avenue for further work. 

Additionally, a further limitation of our work is that the same number of patients per cluster period were assumed, and the designs were all assumed to be cross-sectional.
Both of these assumptions are common at the design stage, however.

Finally, we note that this work could be readily further adapted for other types of outcomes, for example count data.
Alternatively, these methods could be applied to optimal sequence weights for a cluster crossover trial design.

\bibliographystyle{apalike2}
\bibliography{references}

\begin{thebibliography}{}

\bibitem[Atkinson et~al., 2007]{atkinson2007}
Atkinson, A., Donev, A., and Tobias, R. 2007.
\newblock {\em Optimum experimental designs, with SAS}.
\newblock OUP.

\bibitem[Barker et~al., 2017]{barker2017}
Barker, D., D'Este, C., Campbell, M., and McElduff, P. 2017.
\newblock Minimum number of clusters and comparison of analysis methods for cross sectional stepped wedge cluster randomised trials with binary outcomes: A simulation study.
\newblock {\em Trials}, 18:119.

\bibitem[Chaloner and Larntz, 1989]{chaloner1989}
Chaloner, K. and Larntz, K. 1989.
\newblock Optimal bayesian design applied to logistic regression experiments.
\newblock {\em Journal of Statistical Planning and Inference}, 21:191--208.

\bibitem[Dror and Steinberg, 2006]{dror2006}
Dror, H. and Steinberg, D. 2006.
\newblock Robust experimental design for multivariate generalized linear models.
\newblock {\em Technometrics}, 48(4):520--9.

\bibitem[Ford and Westgate, 2020]{ford2020}
Ford, W. and Westgate, P. 2020.
\newblock Maintaining the validity of inference in small‐sample stepped wedge cluster randomized trials with binary outcomes when using generalized estimating equations.
\newblock {\em Statistics in Medicine}, 39:2779--92.

\bibitem[Golden et~al., 2015]{golden2015}
Golden, M., Kerani, R., Stenger, M., Hughes, J., Aubin, M., Malinski, C., and Holmes, K. 2015.
\newblock Uptake and population-level impact of expedited partner therapy (ept) on chlamydia trachomatis and neisseria gonorrhoeae: The washington state community-level randomized trial of ept.
\newblock {\em PLOS Medicine}, 12:e1001777.

\bibitem[Grayling et~al., 2017]{grayling2017}
Grayling, M., Wason, J., and Mander, A. 2017.
\newblock Stepped wedge cluster randomized controlled trial designs: A review of reporting quality and design features.
\newblock {\em Trials}, 18:33.

\bibitem[Harrison and Wang, 2021]{harrison2021}
Harrison, L. and Wang, R. 2021.
\newblock Power calculation for analyses of cross‐sectional stepped‐wedge cluster randomized trials with binary outcomes via generalized estimating equations.
\newblock {\em Statistics in Medicine}, 40:6674--88.

\bibitem[Hussey and Hughes, 2007]{hussey2007}
Hussey, M. and Hughes, J. 2007.
\newblock Design and analysis of stepped wedge cluster randomized trials.
\newblock {\em Contemporary Clinical Trials}, 28:182--91.

\bibitem[Kushner, 1997]{kushner1997}
Kushner, H. 1997.
\newblock Optimal repeated measurements designs: The linear optimality equations.
\newblock {\em The Annals of Statistics}, 25:2328--44.

\bibitem[Kushner, 1998]{kushner1998}
Kushner, H. 1998.
\newblock Optimal and efficient repeated-measurements designs for uncorrelated observations.
\newblock {\em Journal of the American Statistical Association}, 93:1176--87.

\bibitem[Laska et~al., 1983]{laska1983}
Laska, E., Meisner, M., and Kushner, H. 1983.
\newblock Optimal crossover designs in the presence of carryover effects.
\newblock {\em Biometrics}, 39:1087--91.

\bibitem[Lawrie et~al., 2015]{lawrie2015}
Lawrie, J., Carlin, J., and Forbes, A. 2015.
\newblock Optimal stepped wedge designs.
\newblock {\em Statistics \& Probability Letters}, 99:210--14.

\bibitem[Li et~al., 2018a]{li2018b}
Li, F., Turner, E., and Preisser, J. 2018a.
\newblock Optimal allocation of clusters in cohort stepped wedge designs.
\newblock {\em Statistics \& Probability Letters}, 137:257--63.

\bibitem[Li et~al., 2018b]{li2018a}
Li, F., Turner, E., and Preisser, J. 2018b.
\newblock Sample size determination for gee analyses of stepped wedge cluster randomized trials.
\newblock {\em Biometrics}, 74:1450--8.

\bibitem[Li et~al., 2022]{li2022}
Li, F., Yu, H., Rathouz, P., Turner, E., and Preisser, J. 2022.
\newblock Marginal modeling of cluster-period means and intraclass correlations in stepped wedge designs with binary outcomes.
\newblock {\em Biostatistics}, 23:772--88.

\bibitem[Liang and Zeger, 1986]{liang1986}
Liang, K. and Zeger, S. 1986.
\newblock Longitudinal data analysis using generalized linear models.
\newblock {\em Biometrika}, 73:13--22.

\bibitem[Pettersson, 2005]{pettersson2005}
Pettersson, H. 2005.
\newblock Optimal design in average for inference in generalized linear models.
\newblock {\em Statistical Papers}, 46:79--99.

\bibitem[Singh, 2024]{singh2024}
Singh, S. 2024.
\newblock Bayesian optimal stepped wedge design.
\newblock {\em Biometrical Journal}, 66:2300168.

\bibitem[Singh and Mukhopadhyay, 2016]{singh2016}
Singh, S. and Mukhopadhyay, S. 2016.
\newblock Bayesian crossover designs for generalized linear models.
\newblock {\em Computational Statistics \& Data Analysis}, 104:35--50.

\bibitem[Thompson et~al., 2017]{thompson2017}
Thompson, J., Fielding, K., Hargreaves, J., and Copas, A. 2017.
\newblock The optimal design of stepped wedge trials with equal allocation to sequences and a comparison to other trial designs.
\newblock {\em Clinical Trials}, 14:639--47.

\bibitem[Wang et~al., 2021]{wang2021}
Wang, J., Cao, J., Zhang, S., and Ahn, C. 2021.
\newblock Sample size and power analysis for stepped wedge cluster randomised trials with binary outcomes.
\newblock {\em Statistical Theory and Related Fields}, 5:162--9.

\bibitem[Woods and van~de Ven, 2011]{woods2011}
Woods, D. and van~de Ven, P. 2011.
\newblock Blocked designs for experiments with correlated non-normal response.
\newblock {\em Technometrics}, 53(2):173--82.

\bibitem[Zeger et~al., 1988]{zeger1988}
Zeger, S., Liang, K., and Albert, P. 1988.
\newblock Models for longitudinal data: A generalized estimating equation approach.
\newblock {\em Biometrics}, 44:1049--60.

\bibitem[Zhan et~al., 2018]{zhan2018}
Zhan, Z., de~Bock, G., and van~den Heuvel, E. 2018.
\newblock Optimal unidirectional switch designs.
\newblock {\em Statistics in Medicine}, 37:3573--88.

\bibitem[Zhou et~al., 2018]{zhou2018}
Zhou, X., Liao, X., Kunz, L., Normand, S., Wang, M., and Spiegelman, D. 2018.
\newblock A maximum likelihood approach to power calculations for stepped wedge designs of binary outcomes.
\newblock {\em Biostatistics}, 21:102--21.

\end{thebibliography}

\end{document}


\title{\textbf{Supplementary materials to `Optimal Bayesian stepped-wedge cluster randomised trial designs for binary outcome data'}}
\date{}
\maketitle

\begin{figure}
    \centering
    \includegraphics[width=\textwidth]{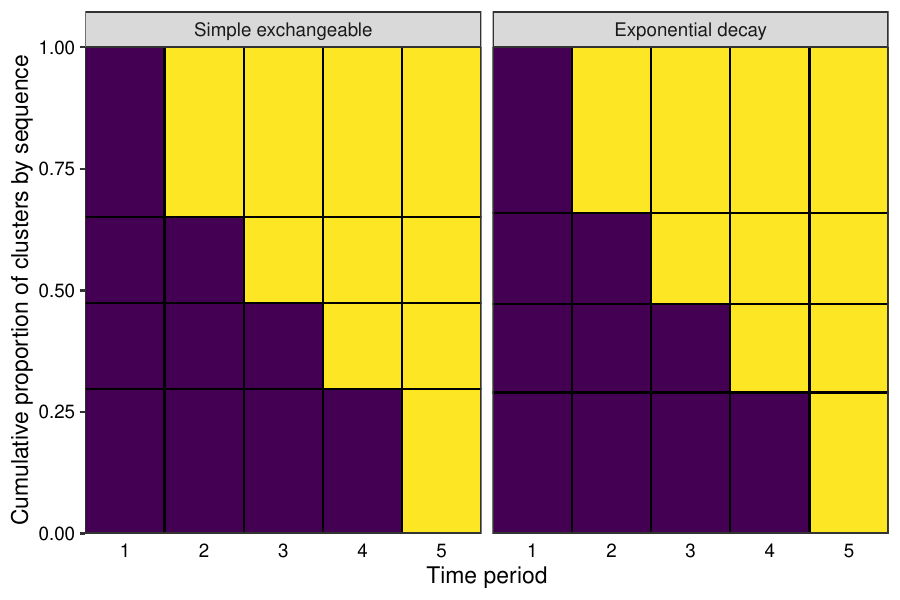}
    \caption{Optimal designs for the Washington State EPT trial example, under the simple exchangeable and exponential decay models.}
\end{figure}

\begin{figure}
    \centering
    \includegraphics[width=\textwidth]{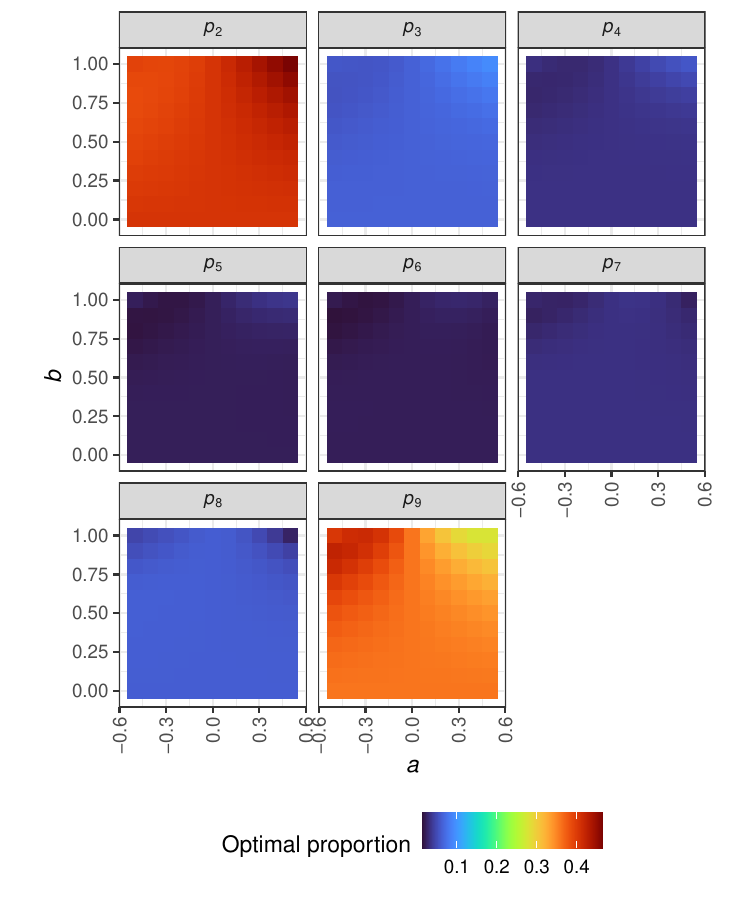}
    \caption{Optimal values of $p_2, \dots, p_9$ in the case where $K = 50$, $\alpha_0 = 0.01$, and $\rho = 0.5$ (exponential decay), as a function of time trend parameters $a$ and $b$.}
\end{figure}

\begin{figure}
    \centering
    \includegraphics[width=\textwidth]{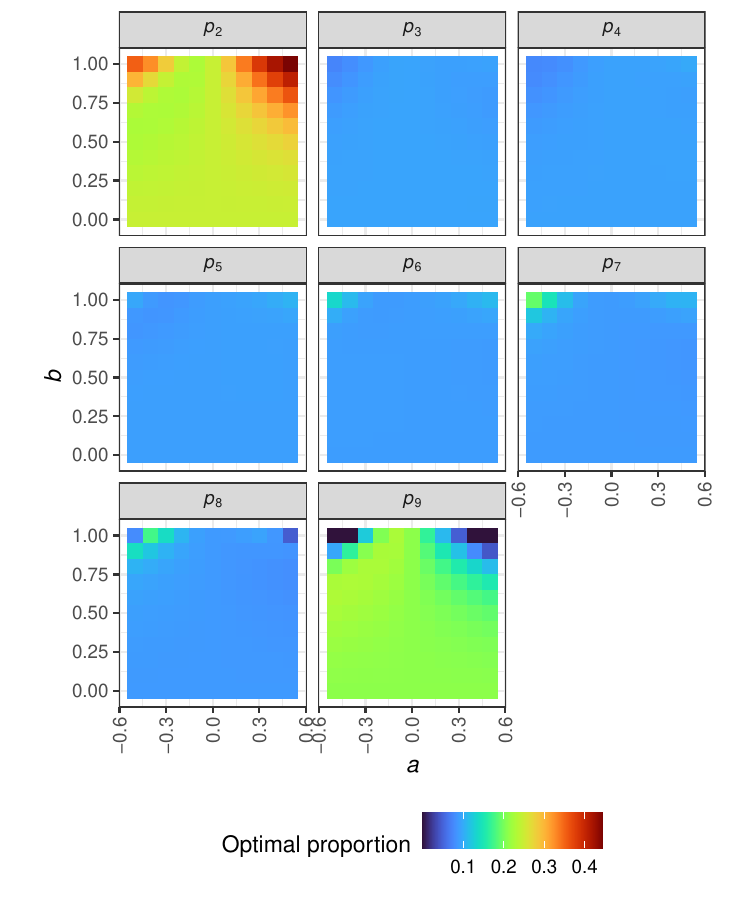}
    \caption{Optimal values of $p_2, \dots, p_9$ in the case where $K = 50$, $\alpha_0 = 0.01$, and $\rho = 1$ (simple exchangeable), as a function of time trend parameters $a$ and $b$.}
\end{figure}